\documentclass[12pt]{article}
\begin{document}
\title{On the Energy Problem in General Relativity\footnote{Published in ``10th
International Conference on General Relativity and Gravitation'',
Padova 4-9 July 1983, Contributed Papers, Ed.: B. Bertotti, F. de
Felice, A. Pascolini. Vol.1, p. 609.}}
\author{Alexander Poltorak\footnote{Current address: General Patent Corporation, Montebello Park, 75 Montebello Road, Suffern, NY 10901-3740 USA. E-mail: apoltorak@gpci.com.}\\Cornell University Medical College, New York, NY}
\date{1983}
\maketitle
\makeatletter
\renewcommand\footnoterule{%
  \vspace{1em}%   <-- one line space between text and footnoterule
  \kern-3\p@\hrule\@width.4\columnwidth%
  \kern2.6\p@}
\makeatother

\begin{abstract}
The Energy Problem (EP) in General Relativity (GR) is analyzed in
the context of GR's axiomatic inconsistencies.  EP is classified
according to its local and global aspects.  The local aspects of
the EP include noncovariance of the energy-momentum pseudotensor
(EMPT) of the gravitational field, non-uniqueness of the EMPT,
asymmetry of EMPT and vanishing metric energy-momentum tensor. The
global aspect of the EP relates to the lack of integral
conservation laws due to the general difficulties in defining
invariant integrals of tensors in non-Euclidean space.  These
difficulties are related to the lack of precise definition of a
reference frame in the GR.  A reference frame is defined here as a
differential manifold with an affine connection.  The resulting
unique decomposition of the Levi-Civita connection into its affine
and nonmetric parts allows for a covariant definition of the
gravitational energy-momentum tensor.  It is pointed out that the
invariance of the Lagrangian (or action functional) is a necessary
but not sufficient condition to secure the covariance of the
Lagrange-Euler field theory.  A rigorous definition of the
Lagrange Field Structure (LFS) on differential manifolds is
proposed.  A covariant generalization of the first Noether theorem
for LFS is obtained.  Different approaches to the EP are
discussed.

\end{abstract}

\begin{enumerate}
\item The EP has different aspects related to the local
(differential) and global (integral) conservation laws.  The local
aspects are:

    \begin{enumerate}
    \item Noncovariance of the energy-momentum pseudotensor (EMPT)
of a gravitational field both as a mathematics problem in itself
and, in its physical setting, as the Lorentz-Bauer-Schr{\"o}dinger
paradox.
    \item Non-uniqueness of the EMPT --- there are the
EMPTs of Einstein, Lorentz-M{\o}ller, Landau-Lifshitz (e.g. [2])
and others.
    \item Asymmetry of most of these EMPTs as well as of the
canonical EMPT; the only symmetric one [3] is not even a
pseudotensor.
    \item The metric energy-momentum tensor (EMT) which
is always real and symmetric is identically equal to zero by
virtue of the field equations.  (The canonical EMPT –-- which
differs from the symmetric one by the covariant divergence of the
spin tensor –-- is also equal to zero [2, 4].)
    \end{enumerate}

The global aspect is that an invariant integral of any symmetrical
2-tensor in non-Euclidian space does not exist (the noninvariant
integration generally used in GR is not valid despite the fact
that it gives correct results in a few special cases) i.e., the
global conservation laws depend on spatial symmetry (because one
can define invariant integration only in symmetrical spaces).
Noether Theory, which is the most consistent approach for
considering conservation characteristics of the field, cannot be
applied in its classical version because of its noncovariance.
Therefore, we do not have a covariant, unique and conservative EMT
for a gravitational field in GR.  But even if we would have one,
we would still not have global (integral) invariant quantities for
an asymmetrical field.

\item Bianchi's identity \( D_\upsilon  T_\mu ^\upsilon   = 0 \),
where \( T_\mu ^\upsilon \) is the EMT of matter and \(D_\upsilon
\) is the covariant derivative with respect to Riemannian
connection \(\Gamma _{\mu \upsilon }^\tau \), can be rewritten in
the form
\begin{equation}
        \partial _\upsilon  \left( {{\rm T}_\mu ^\upsilon   + t_\mu ^\upsilon  } \right)
\end{equation}
where ${t_\mu ^\upsilon  }$ is the EMPT of the gravitational field
and $\partial _\upsilon  $ is a partial derivative.  Because of
the way we defined ${t_\mu ^\upsilon  }$, it is not a tensor – the
procedure of picking out a partial derivative from a covariant one
is not invariant.   By using a superpotential \( \Psi _\mu
^{\lambda \upsilon } \), $t_\mu ^\upsilon   = \partial _\tau \Psi
_\mu ^{\tau \upsilon } $, $t_\mu ^\upsilon  $ is defined
noncovariantly from the very beginning. In the same way, the
canonical EMPT is not a tensor because to form it we take a
partial (noncovariant) derivative of the Lagrangian.

The geometrical structure of GR is $\left\langle {M,g,\Gamma }
\right\rangle $, where \emph{M} is a differential manifold,
\textbf{g} is a metric on \emph{M} and $\Gamma $ is a Riemannian
connection on \emph{M} such that $D_\tau  g_{\mu \upsilon }  = 0$.
 The metric $g_{\mu \upsilon } $ cannot serve as a potential of any field at all; if it did we
would have a metrical EMT equal to zero which is unacceptable (it
means that there is no field).  Nor can the connection $\Gamma
_{\mu \upsilon }^\tau  $ be a potential of a field because it is
not a covariant object and includes some extra information about
the coordinate system.  So, if one takes $\Gamma _{\mu \upsilon
}^\tau  $ as a potential of the gravitational field, the field is
nonlocalized by definition.

The analysis of the integral conservation laws, which are valid
only in symmetrical spaces, leads to the same conclusions --– if
the gravitational field is described by the metric $g_{\mu
\upsilon } $ of Riemannian space and symmetry of the space
reflects the symmetry of the field, then the conservation laws
exist only for symmetrical fields.

Another serious problem in GR is the definition of reference
frames.  Since they are not defined in GR, everyone's
understanding of them is something different.  We should mention
that there is no justification to think that reference frame and
coordinate system are one and the same.  It has been proven and
illustrated in many works (e.g.[5]) that a coordinate system is
strictly a formal mathematical object, which has no physical, and
very little mathematical, meaning, just as a reference frame is a
real physical object.

In any case, we take as an axiom (because the truth of it is
obvious to us) that a reference frame is a differentiable manifold
with an affine connection $\nabla $, which also plays the role of
the inertia field potential.  As we saw, the geometrical structure
$\left\langle {M,g,\Gamma } \right\rangle $ seems to leave no room
for both the reference frame and the gravitational field.

Fortunately, it is not quite so.  For each Riemannian connection
and metric \emph{g}, there exists an affine connection $\nabla $
such that $\Gamma  = \nabla  + S$, where \emph{S} is the nonmetric
part of $\nabla $ (called tensor of nonmetricity).  So the
geometrical structure of GR is actually $\left\langle {M,g,\nabla
,S} \right\rangle $. As soon as we assume that $\nabla $ is an
inertia field potential, it can easily be shown within the
framework of GR that \emph{S} is a gravitational field itself [6].
This solves the problem immediately.  Instead of using partial
derivatives, we should use covariant derivatives $\nabla _\mu  $
with respect to the affine connection $\nabla $. If we now define
an EMT of a gravitational field as $t_\mu ^\upsilon   = \nabla
_\tau  \Psi _\mu ^{\tau \upsilon } $, where $\Psi _\mu ^{\tau
\upsilon } $ is a tensor ``superpotential,'' it gives us covariant
differential conservation laws:
\begin{equation}
\nabla _\upsilon  \left( {T_\mu ^\upsilon   + t_\mu ^\upsilon  }
\right) = \frac{1} {2}\left\{ {\nabla _\upsilon  ,\nabla _\tau  }
\right\}\Psi _\mu ^{\tau \upsilon }
\end{equation}

Existence of integral conservation laws does not depend on the
symmetry of a gravitational field, but on the symmetry of the
chosen reference frame.  So, in the inertial reference frame there
are always ten covariant conservative quantities.

\item  As has been noted previously [6], the Lagrange formalism
generally used in GR is not rigorous enough.  Really, for the
Lagrangian L=L$(\phi ,\partial _\mu  \phi , \ldots )$ (where $\phi
$ is some field), which is covariant but depends on noncovariant
arguments, the "canonical momentum" of the field $\pi ^\mu   =
L/\partial \left( {\partial _\mu  \phi } \right)$ turns out to be
noncovariant (for simplicity, only first partials are indicated).
This immediately leads to noncovariance of the canonical EMT
$T_\mu ^\upsilon   = \pi ^\upsilon  \partial _\mu  \phi  - \delta
_\mu ^\upsilon  L$, where $\delta _\mu ^\upsilon  $ is the
Kronecker tensor.  This situation occurs in GR, where the Hilbert
Lagrangian $L = R\left( {g_{\mu \upsilon } ,\partial _\tau  g_{\mu
\upsilon ,} \partial _\sigma  \partial _\tau  g_{\mu \upsilon } }
\right)$ (\emph{R} is the Riemannian curvature scalar) results in
noncovariance of the EMT of the gravitational field.

Thus, invariance of the Lagrangian, or more rigorously, invariance
of the action functional is not a sufficient condition to secure
the covariance of the theory, but it has mistakenly been thought
of as such.  This misunderstanding has come about due to the
absence of a rigorous definition of an invariant Lagrange field
structure on a manifold.  Here we shall provide this structure.
Let \emph{M} be a differentiable manifold with affine connection
$\nabla $ and metric \emph{g}. Then let $\phi $ be a
differentiable tensor (or spinor) field on \emph{M} and $J_k
\left( \phi  \right)$ be the \emph{k}-th order jet bundle over
\emph{M}. Let us construct a morphism $L:J_k \left( \phi  \right)
\to R$.

The triplet $\Lambda  = \left\langle {M,\phi ,L} \right\rangle $
will be called a Lagrange Field Structure (LFS) on the
differentiable manifold with Lagrangian \emph{L}. Let $\xi $ be a
chart with field of definition \emph{U}, then $L = L\left( {\phi
,\nabla _\mu  \phi ,\nabla _\upsilon  \nabla _\mu  \phi ,...}
\right)$, where ${\nabla _\mu  }$ is the covariant derivative with
respect to affine connection $\nabla $.  So in GR we have to write
$L = R\left( {g_{\mu \upsilon } ,\nabla _\tau  g_{\mu \upsilon }
,\nabla _\sigma  \nabla _\tau  g_{\mu \upsilon } } \right)$,
$\left( {\nabla _\tau  g_{\mu \upsilon }  \ne 0} \right)$ or $L =
R\left( {g_{\mu \upsilon } ,S_{\mu \upsilon }^\tau  ,\nabla
_\sigma  S_{\mu \upsilon }^\tau  } \right)$.

Now we have $\pi  = L/\partial \left( {\partial _\upsilon  \phi }
\right)$, $T_\mu ^\upsilon   = \pi ^\upsilon  \partial _{\mu \phi
}  - \delta _\mu ^\upsilon  L$ and
\begin{equation}
\nabla _\mu  T_\upsilon ^\mu   = \pi ^\tau  \left\{ {\nabla _\tau
,\nabla _\upsilon  } \right\}\phi
\end{equation}

As we mentioned above, the Noether theory cannot be applied to GR
in its noncovariant form.  Due to this, we provided covariant
generalization of the first Noether theorem for GR, based on the
invariant definition of LFS. As a result of it, the following
theorem has been obtained:

Let $\Lambda  = \left\langle {M,\phi ,L} \right\rangle $ be a LFS.
If action functional $S\left[ \phi  \right]$ is invariant under a
finite r-parametrical group Gr, then the \emph{r }linearly
independent combinations of Lagrange derivatives become
divergences up to an additive geometry-dependent term, which plays
a role similar to the additional sources in a nonhomogeneous
Euler-Lagrange equation.

\item Generally speaking, three points of view on EP are possible:
(1) The EP is not a problem at all, but rather the specific
property of GR related to the Principle of Equivalence [7]; (2)
The EP is a fatal problem of the theory and GR must be replaced by
another theory free of the EP (like different versions of
bimetrism, e.g. [8,9]); and (3) The EP is a serious problem,
rooted in some incorrectness of GR, but which should be solved
within the framework of GR.

The first point of view is in error because it is based on the
incorrect assumption that reference frame and coordinate system
are one and the same.  The second point of view would be right if
we would not be able to solve the problem within the framework of
GR.  Since we have shown here a few possible ways to solve the
problem, there is no reason to reject GR.

In fact, we did not modify GR, but resolved intrinsic
contradictions based on GR principles.  These corrections result
in a theory which differs minimally from GR and which is not an
alternative, but rather the correct form of GR.
\end{enumerate}

\large \textbf{References}
\normalsize
\begin{enumerate}
\item Lorentz, H.A.  Veral. Kön. Akad. Wet., Amsterdam, 25, 1380,
1916.
\item Einstein, A. Sitz. Preuss. Akad. Wiss., 1, 154, 1918.
\item Landau, L.,
Lifshitz, E.  Field Theory, Moscow, 1962.
\item Einstein, A. Sitz.
Preuss. Akad. Wiss., 2, 11, 1916.
\item Rodichev, V.I.  Theory of
Gravitation in an Orthogonal Frame, Nauka, Moscow, 1974.
\item Poltorak, A.  GR9 Conference Abstracts, Jena, 2, 516, 1980.
(Available on www.arXiv.org as gr-qc/0403050)
\item Misner, C., Thorne, K., Wheeler, J.  Gravitation, Freeman, San
Francisco, 1973, p. 466.
\item Rosen, N.  Phys. Rev., 57, 147,
1940.
\item Logunov, A., Folomeshkin, V.  Theor. Math. Phys.,
Moscow, 32, 2, 174, 1977.
\end{enumerate}

\end{document}